\documentclass{emulateapj}

\slugcomment{Accepted manuscript}
\shorttitle{Spitzer Spectrospcopy of NGC\,7714}
\shortauthors{Brandl et al.}

\begin{document}

\title{Spitzer-IRS Spectroscopy of the Prototypical Starburst Galaxy
       NGC\,7714}


\author{B.R. Brandl}
\affil{Leiden University, P.O. Box 9513, 2300 RA Leiden, The Netherlands}
\email{brandl@strw.leidenuniv.nl}
\author{D. Devost, S.J.U. Higdon,
        V. Charmandaris\altaffilmark{\ddag}, D. Weedman,
        H.W.W. Spoon, T.L. Herter, L. Hao, J.~Bernard-Salas,
        J.R.~Houck} \affil{Cornell University, Astronomy Department,
        Ithaca, NY 14853}
\altaffiltext{\ddag}{Chercheur Associ\'e, Observatoire de Paris, F-75014,
        Paris, France}
\and
\author{L. Armus, B.T. Soifer, C.J. Grillmair, P.N. Appleton}
\affil{Caltech Spitzer Science Center, 314-6, Pasadena, CA 91125}

\begin{abstract}
We present observations of the starburst galaxy NGC\,7714 with the
Infrared Spectrograph {\sl IRS}\footnote{The {\sl IRS} was a
collaborative venture between Cornell University and Ball Aerospace
Corporation funded by NASA through the Jet Propulsion Laboratory and
the Ames Research Center.} on board the {\sl Spitzer Space Telescope}.
The spectra yield a wealth of ionic and molecular features that allow
a detailed characterization of its properties. NGC\,7714 has an H\,II
region-like spectrum with strong PAH emission features.  We find no
evidence for an obscured active galactic nucleus, and with
[Ne\,III]\,/\,[Ne II]\,$\approx 0.73$, NGC\,7714 lies near the upper
end of normal-metallicity starburst galaxies.  With very little
slicate absorption and a temperature of the hottest dust component of
340\,K, NGC\,7714 is the perfect template for a young, unobscured
starburst.
\end{abstract}



\section{Introduction}

NGC\,7714 is a peculiar barred-spiral galaxy at $45^{\circ}$
inclination.  For $cz = 2798$\,km/s the galaxy is at
distance\footnote{We adopt $H_{\rm 0} = 71$\,km\,s$^{-1}$ Mpc$^{-1},
\Omega_{\rm M}=0.27, \Omega_{\rm \Lambda}=0.73$} of 39.6\,Mpc .
Together with its post-starburst companion NGC\,7715 it forms the
interacting system Arp~284, which is the result of a recent
($100-200$\,Myr), off-center collission between the two disk galaxies
\citep{str03}.  With its compact, UV-bright nucleus, NGC\,7714 has
been classified by \citet{wee81} as the proto-typical starburst
galaxy.  The central region of about 330\,pc has been the site of
active star formation at a rate of about $1 M_\sun$\,yr$^{-1}$ for
some $10^8$~years.  However, a recent significant increase in the star
formation rate made it the dominating source of the UV flux
\citep{lan01}.

With its strong He{\sc II} $\lambda$4686 line \citep{gon95}, NGC\,7714
has been classified as a Wolf-Rayet galaxy.  In fact, the optical and
UV spectra indicate a population of about 2000 Wolf-Rayet and 20000
O-type stars, suggesting a fairly young age of the present starburst of
$4-5$~Myr \citep{gar97}.  This is in agreement with earlier studies by
\citet{wee81} and \citet{tan88} who estimated $~10^4$~O5-type stars
from the Balmer and Bracket-$\gamma$ line fluxes, and masses of
ionized gas of $3.0\times 10^6 M_\sun$, and $1.9\times 10^6 M_\sun$,
respectively. The radio-continuum and the X-ray luminosity of $6\times
10^{40}$~ergs\,s$^{-1}$ require $~10^4$~supernova remnants in a volume
of radius 280\,pc \citep{wee81}. \citet{tan88} found definite evidence
for a starburst-driven bipolar winds from the nucleus, nearly
perpendicular to the disk plane.

Although most of the activity is concentrated in the nucleus,
NGC\,7714 as a whole is experiencing intense star formation
\citep{gon95}.  Mid-IR imaging with {\sl ISOCAM} \citep{oha00}
revealed a strong source at the nucleus surrounded by slightly
extended emission out to about $30''$ (5.7~kpc) in diameter.  From the
{\sl IRAS} fluxes \citep{sur04} and the above distance we calculate a
total infrared luminosity of $L_{8-1000\mu m} = 5.6\times 10^{10}
L_\sun$.  

The strong optical emission lines, H$\beta$ and [O\,III], show no
signs of broad emission (\citet{wee81}, \citet{tan88}).  Recently,
\citet{sor04} found two compact X-ray sources with {\sl XMM-Newton}.
One of them coincides with the starburst nucleus, has an X-ray
luminosity of $L_X\approx 10^{41}$~erg\,s$^{-1}$, and shows the
spectrum of a thin thermal plasma with a power-law (point-source)
contribution.  The variability in the power-law component hints at the
presence of either a hidden low luminosity AGN or an ultraluminous
X-ray source \citep{sor04}.

In this letter we describe new observations of NGC\,7714 with the {\sl
Spitzer Space Telescope}.  These observations are part of an IRS
guaranteed time program to obtain high signal-to-noise spectra of a
large sample of different classes of nearby galaxies, that can be used
for comparison with more distant systems (Devost et al. 2004, Higdon
et al. 2004a, Houck et al. 2004a).


\section{Observations and Data Reduction}
\label{secobservations}
\label{secdatared}

We observed NGC\,7714 with the Infrared Spectrograph ({\sl IRS})
\citep{hou04b} on board the {\sl Spitzer Space Telescope}
\citep{wer04}.  The observations are part of the {\sl IRS} guaranteed
time program.  The data were taken during the first
{\sl IRS} campaign in nominal operations on 17 December 2003 using the
standard IRS ``Staring Mode'' Astronomical Observing Template (AOT).
The dataset consists of observations with all four IRS modules:
2 cycles $\times$ 14s in ``Short-low'' 
(SL, $\Delta \lambda = 5.2 - 14.5\mu$m, $R \sim 64 - 128$),
2 cycles $\times$ 14s in ``Long-low''
(LL, $\Delta \lambda = 14.0 - 38.0\mu$m, $R \sim 64 - 128$), 
4 cycles $\times$ 30s in ``Short-high''
(SH, $\Delta \lambda = 9.9 - 19.6\mu$m, $R \sim 600$), and 
2 cycles $\times$ 60s in ``Long-high''
(LH, $\Delta \lambda = 18.7 - 37.2\mu$m, $R \sim 600$).
Each cycle yields two exposures at different nod positions along the
slit.  The slits were positioned relative to the reference star
SAO~128273 using an ``IRS high accuracy peak-up'' in the blue filter 
band.

The data have been pre-processed by the Spitzer Science Center (SSC)
data reduction pipeline version 9.5 (Spitzer Observer's Manual,
chapter~7\footnote{http://ssc.spitzer.caltech.edu/documents/som/}).
The two-dimensional ``basic calibrated data'' (BCD) constituted the
basis for further processing.  The BCD frames have been individually
inspected by eye, and hot pixels, as well as very negative pixels,
that have not already been flagged by the pipeline, were masked
manually.  The former could easily be identified by flipping between
images at different nod positions.  Tests have shown that masked
pixels introduce artificially high noise in the spectra, and hence we
interpolated isolated masked pixels by their nearest neighbor pixel
values along the dispersion direction.  Next we averaged the frames
from the same nod positions (in the case of SH where four frames per
nod position were available we have used the median instead).  Since
the low-resolution, long-slit modules contain two sub-slits, each
integration provides a ``free'' sky spectrum -- mainly zodiacal
emission -- in the adjacent subslit.  We computed the median sky and
subtracted it from the low-resolution frames.

Further processing was done within the {\sl IRS} data reduction and
analysis package {\sl SMART}, version v.4 \citep{hig04b} -- a powerful
IDL package for spectral extraction and spectral analysis.  The
high-resolution spectra were extracted with ``full aperture"
extraction along the diffraction orders.  The ends of each orders
where the noise increases significantly were manually clipped.  Within
any one module the individual orders matched remarkably well and required
no further fine-tuning, with the exception of SH order 11 which was
scaled up by 5\%.  Finally the two nod positions were averaged.  The
low-resolution spectra were extracted using {\sl SMART}'s
``interactive column extraction'', which is similar to the method used
in the SSC pipeline.  The 3rd ``bonus'' order in SL and LL has not
been included.

\notetoeditor{Place figure~\ref{fighires} here, if possible on the same page
              and side by side with figure~\ref{figlores}.}

\notetoeditor{Place figure~\ref{figlores} here, if possible on the same page
              and side by side with figure~\ref{fighires}.}

At the current stage of calibration, there remains a significant
mismatch between some modules in both low- and high-resolution
spectra.  This mismatch is most likely due to either {\sl (i)}
extended source emission (as seen e.g. by \citet{oha00}), or {\sl
(ii)} pointing errors that lead to flux losses that are most
significant at short wavelengths (narrow slits).  We also note that
the low-resolution slits are narrower than the slits of the
high-resolution modules.  At this early stage of the Spitzer mission
we have no means to identify the real cause.  However, the LH and LL
spectra agree very well with each other, and a comparison between the
{\sl IRAS} $25\mu$m flux of 3.15~Jy \citep{sur04} and the LL
flux in the {\sl IRAS} filterband of 2.61~Jy yields good agreement,
taking into account that the {\sl IRAS} ``aperture'' is much bigger
and is even more susceptible to extended emission.  Adopting the LH
flux densities as correct, we scaled SH up by 36\% to achieve an
excellent match of the continuum fluxes around $19\mu$m.  We note that
the flux in the [S\,III] line, which lies in the overlap region
between SH and LH, is then the same in both modules.  Similarly, the
low-resolution spectra from different modules had to be scaled to
match.  The LL 2nd order was scaled up by 17\% to match the LL 1st
order; the SL 2nd order was scaled up by 40\% to match the SL 1st
order; and finally the resulting SL spectrum was scaled up by 17\% to
match the LL spectrum.  As a result of this approach the low- and high
resolution spectra agree very well with each other, and -- even more
important -- we determine from the low-resolution spectrum in the {\sl
IRAS} $12\mu$m band a total flux of 0.47~Jy, while \citet{sur04}
found 0.56~Jy from {\sl IRAS} -- again good agreement!
Figures~\ref{fighires} and \ref{figlores} show the final {\sl IRS}
high- and low-resolution spectra, respectively.

The properties of the ionic and molecular features were obtained from
single Gaussian fits to the high- and low- resolution spectra within
{\sl SMART}; the results are listed in tables~\ref{tabfine}
and~\ref{tabpah}.


\section{Results} 
\label{secresults}

\subsection{The starburst properties of NGC\,7714}

Based on optical and near-IR spectroscopy \citep{lan01}, the
{\sl ISOCAM} filter band ratios and PAH emission features
\citep{oha00}, it has long been argued that the nucleus of NGC\,7714
is a site of intense starburst activity.  However, given its low
extinction, the unusually high infrared luminosity and recent reports
of a possible, hidden AGN \citep{sor04} more precise diagnostics are
needed to understand the processes in the nuclear region.  The high
signal-to-noise mid-IR {\sl IRS} spectra provide the ideal diagnostic
tools to characterize the nature of the underlying power source in
more detail.

\begin{figure}
\epsscale{1.2}
\plotone{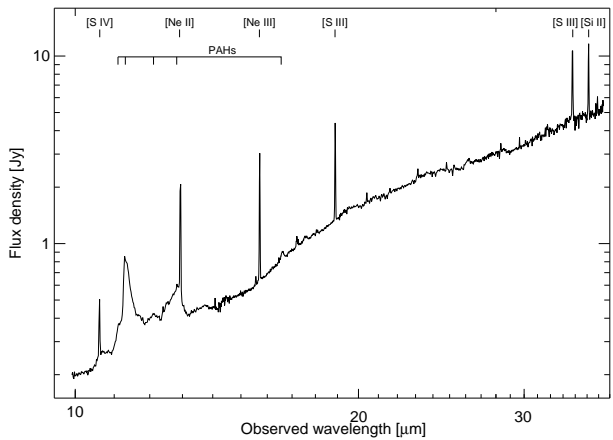}
\caption{IRS high-resolution spectrum of NGC\,7714.  The total integration 
         times are 240~seconds for both SH and LH (see discussion in the 
         main text). \label{fighires}}
\end{figure}

Table~\ref{tabfine} lists the properties of the strongest
fine-structure lines in the high-resolution spectrum
(fig.~\ref{fighires}).  We find $10.8\times 10^{-20}$Wcm$^{-2}$ for
the [Ne\,II] line while \citet{phi84} found $(7\pm 2)\times
10^{-20}$Wcm$^{-2}$ from the ground.  For the [S IV] line we determine
$1.8\times 10^{-20}$Wcm$^{-2}$ while \citet{oha00} quote $(3.2\pm
0.8)\times 10^{-20}$\,Wcm$^{-2}$ based on {\sl ISOPHOT-S}
low-resolution spectra. The S(0), S(1), and S(2) lines of molecular
hydrogen are marginally detected and will be discussed in a subsequent
paper.

Of particular interest is the [Ne\,III]\,/\,[Ne\,II] ratio as it is a
good measure of the hardness of the interstellar radiation field,
which is mainly determined by the most massive stars.  We find
[Ne\,III]\,/\,[Ne II]\,=\,$0.73\pm 0.05$, corresponding to an
effective temperature of about $3.8\times 10^4$\,K \citep{giv02}.
Based on the ``{\sl ISO-SWS} Starburst Sample'', \citet{tho00} studied
the [Ne\,III]\,/\,[Ne\,II] ratio for 27 galaxies, with values ranging
from 0.05 (M83) to 12.0 (II\,Zw\,40), and a median value of 0.26.  The
starbursts of the nearby ISO sample with [Ne III]\,/\,[Ne\,II] ratios
closest to our value are NGC\,3690 (0.71) and the interaction region
in NGC\,4038/39 (0.84).  The origin of these variations is still
controversial (Rigby \& Rieke 2004, Thornley et al. 2000).  Even
within the same galaxy the [Ne III]\,/\,[Ne\,II] ratio can vary
significantly as recently seen with the {\sl IRS} in NGC\,253
\citep{dev04}.  We will address this important issue in a subsequent
paper once we have observed a larger sample of starburst galaxies.
However, with its large number of very luminous Wolf-Rayet stars, it
is not surprising that the [Ne\,III]\,/\,[Ne\,II] ratio in NGC\,7714
lies above the value found in most starburst galaxies.

\begin{deluxetable}{cllrrrc}
\tabletypesize{\scriptsize}
\tablecaption{Fine structure lines in NGC\,7714}
\tablewidth{0pt}
\tablehead{
\colhead{ID} & \colhead{$\lambda_{\mbox{rest}}$} & \colhead{$\lambda_{\mbox{obs}}$} & \colhead{EP\tablenotemark{a}} & \colhead{Flux\tablenotemark{b}} & \colhead{S/N\tablenotemark{a}} & \colhead{EW\tablenotemark{a}}\\
\colhead{} & \colhead{[$\mu$m]} & \colhead{[$\mu$m]} & \colhead{[eV]} & \colhead{[$10^{-20}$Wcm$^{-2}$]} &  \colhead{} & \colhead{[$\mu$m]}
}
\startdata
\mbox{[S IV]}   & 10.51 & 10.61 & 34.8 &  $1.79 \pm 0.04$ & 92  & 0.03\\ 
\mbox{[Ne\,II]}  & 12.81 & 12.94 & 21.6 & $10.82 \pm 0.48$ & 135 & 0.11\\ 
\mbox{[Ne\,III]} & 15.56 & 15.63 & 41.0 &  $7.95 \pm 0.11$ & 114 & 0.11\\
\mbox{[S\,III]}  & 18.71 & 18.89 & 23.3 &  $9.01 \pm 0.10$ & 131 & 0.08\\ 
\mbox{[S\,III]} & 33.48 & 33.77 & 23.3 & $12.61 \pm 0.51$ & 37 & 0.11\\ 
\mbox{[Si\,II]} & 34.82 & 35.13 & 8.15 & $10.45 \pm 0.36$ & 48 & 0.09
\enddata
\label{tabfine}
\tablecomments{Emission line properties obtained from a single Gaussian fit 
               to the high-resolution data.}
\tablenotetext{a}{EP = Excitation potential, EW = Equivalent width (observed), 
                  S/N = signal-to-noise ratio.}
\tablenotetext{b}{The uncertainties quoted for the line fluxes throughout 
                  this letter are the errors from the line fit and do not 
                  include the calibration uncertainties.} 
\end{deluxetable}

Although [O\,IV] has an excitation potential of 54.9\,eV, faint
emission has been seen in many starburst galaxies \citep{lut98} and
recently with the {\sl IRS} in the ultra-luminous IR-galaxy UGC\,5101
\citep{arm04}.  We also find a marginal detection of the [O\,IV] line
with $(9.9\pm 7.2)\times 10^{-21}$\,Wcm$^{-2}$.  Since the location of
the line coincides with a noisy pixel, we take this value as an upper
limit.  Hence, with [O\,IV]\,/\,[Ne\,II]~$\leq 0.09$ and a relative
PAH($7.7\mu$m)\,/\,continuum($7.7\mu$m)~$\approx 1.30$, NGC\,7714
falls below the region occupied by ULIRGs in fig.~5 of \citet{gen98}.
Furthermore, the absence of the [Ne\,V] line -- we place a $3\sigma$
upper limit of $3.9\times 10^{-21}$\,Wcm$^{-2}$ -- emphasizes the pure
starburst nature of NGC\,7714.  On the basis of the infrared emission
lines, therefore, we find no evidence for an obscured ionized region
associated with an AGN, which implies that the variable X-ray source
reported by \citet{sor04} is either from strong shocks associated with
very recent supernova activity, or emission from a HMXB. There is
growing evidence that both possibilities are intimately associated
with massive young star clusters \citep{gao03}.

\subsection{The dominant PAH features}
\label{secpahs}

\begin{figure}
\epsscale{1.2}
\plotone{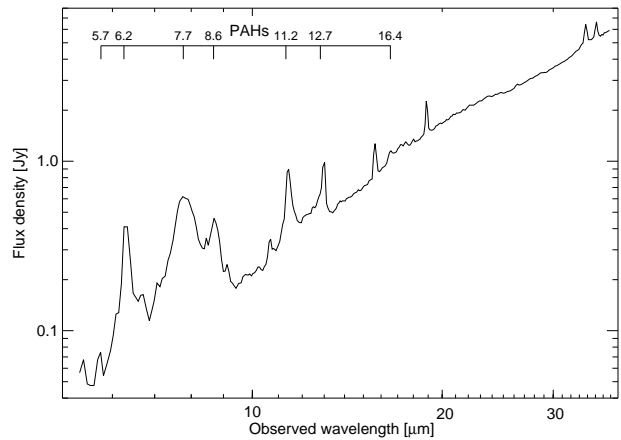}
\caption{IRS low-resolution spectrum of NGC\,7714 (see discussion in
         the main text).  The total integration time is 56~seconds per
         subslit.\label{figlores}}
\end{figure}

\begin{deluxetable}{rrrccc}
\tabletypesize{\scriptsize}
\tablecaption{PAH emission features in NGC\,7714}
\tablewidth{0pt}
\tablehead{
 \colhead{$\lambda_{\mbox{rest}}$} & \colhead{$\lambda_{\mbox{obs}}$} & \colhead{Flux} & 
\colhead{EW} & \colhead{Relative} & \colhead{Mod\tablenotemark{a}}\\
\colhead{[$\mu$m]} & \colhead{[$\mu$m]} & \colhead{[$10^{-19}$Wcm$^{-2}$]} & 
\colhead{[$\mu$m]} & \colhead{strength\tablenotemark{b}} & 
}
\startdata
 6.2 &  6.31 & $4.60 \pm 0.44$ & 0.50 & 0.49 & lo\\
 7.7 &  7.80 & $9.42 \pm 0.39$ & 0.70 & 1.00 & lo\\
 8.6 &  8.71 & $1.74 \pm 0.28$ & 0.17 & 0.18 & lo\\
11.2 & 11.33 & $2.21 \pm 0.28$ & 0.17 & 0.23 & hi\\
12.7 & 12.81 & $1.66 \pm 0.16$ & 0.14 & 0.18 & hi\\
16.4 & 16.55 & $0.33 \pm 0.03$ & 0.03 & 0.04 & lo
\enddata
\label{tabpah}
\tablecomments{PAH band emission strength, For the measurement of the
               $12.7\mu$m feature we first removed the [Ne\,II] line.}
\tablenotetext{a}{lo = fit to low-resolution data, hi = fit to 
                  high-resolution data}
\tablenotetext{b}{Relative to the $7.7\mu$m feature.}
\end{deluxetable}

Below $13\mu$m, the spectra are dominated by strong emission from PAHs
(see e.g. \citet{pee03} for a recent overview), with the strongest at
$6.2\mu$m, $7.7\mu$m, $8.6\mu$m, $11.2\mu$m, $12.7\mu$m, and
$16.4\mu$m listed in table~\ref{tabpah}.  More features (e.g. at
$5.7\mu$m) appear to be present but will not be discussed here.  Some
of those features have also been seen in the {\sl ISOPHOT-S} spectra
which covered the $5-12\mu$m range at a resolution of $R=90$
\citep{oha00}.  The total flux we determine for the strongest feature
at $7.7\mu$m agrees with the {\sl ISOPHOT-S} measurement of $(8.84\pm
1.10)\times 10^{-19}$\,Wcm$^{-2}$.

Shortward of $13\mu$m the spectrum is qualitatively very similar to
the well-know starburst galaxy M82.  However, unlike M82, which has
very strong PAH emission with respect to the warm continuum from very
small grains and plausibly some extinction due to silicates at
$9.7\mu$m \citep{for03}, the spectrum of NGC\,7714 shows a rather 
smooth continuum and no evidence for silicate absorption. 

With the high signal-to-noise provided by {\sl Spitzer-IRS} it now
becomes possible to expand the comprehensive study of fainter PAH
features seen in Galactic sources to a large sample of extragalactic
sources.  While the strength of e.g. the $6.2\mu$m PAH / continuum
ratio is remarkably constant over a wide luminosity range in Galactic
sources, others, e.g. the $11.2\mu$m feature, which is linked to
neutral PAHs, get quickly reduced in stronger ionizing radiation
fields \citep{hon01}.

\citet{hon01} show that the $12.7\mu$m band correlates well with the
CC stretch mode at $6.2\mu$m for Galactic sources ranging from YSOs,
H\,II regions, and reflection nebulae to planetary nebulae (their
fig.~5).  With $\log_{10}(I_{6.2} / I_{11.2}) = 0.32$ and
$\log_{10}(I_{12.7} / I_{11.2}) = -0.12$, NGC\,7714 falls right on
their correlation line.  We conclude that the physical conditions in
NGC\,7714 are similar to Galactic H\,II regions.  In fact, a
comparison with the {\sl ISO-SWS} spectrum of M17 \citep{pee04} yields
a perfect overall match longward of $15\mu$m.

\subsection{Dust and Extinction in NGC\,7714}
\label{secdust}

One of the most remarkable attributes in Figure~\ref{figlores} is the
lack of silicate absorption in NGC\,7714.  This implies that the
observed spectral shape of the mid-IR continuum is defined purely by
the continuum emission of the hottest dust directly heated by the
young, massive stars. If so then NGC\,7714 might be a well-needed
template for fitting relatively unobscured starbursts at low and high
redshifts.  Is NGC\,7714 the perfect, unobscured starburst?

From the hydrogen recombination lines Pa$\beta$, Br$\gamma$, and
Br$\alpha$, \citet{pux94} determined $A_V = 1.8\pm 0.7$~mag for a point
source model with obscuring screen, and $A_V = 3.9\pm 1.7$~mag if
sources and dust are homogeneously mixed.  For an extinction law of
$A_{9.6\mu m} / A_V = 0.1$ and if $A_V = 3.9$ we find that
$\tau_{9.6\mu m} = 0.36$ and hence the continuum around $9.6\mu$m
would be suppressed by a factor of at most 30\%.  In the case of an
obscuring screen, which was favored by \citet{pux94}, the suppression
is only 15\%, which is in good agreement with the shape of our
spectrum.

\begin{figure}
\epsscale{1.2}
\plotone{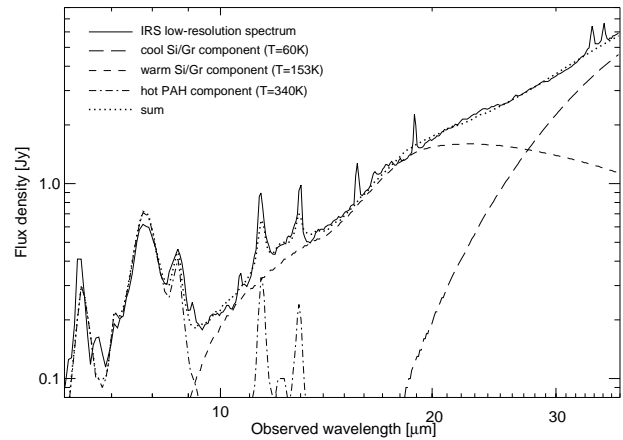}
\caption{IRS low-resolution spectrum of NGC\,7714 with the various fit 
         components overlaid (see discussion in the text).  Strong 
         emission lines were excluded from the fit. \label{figbbfit}}
\end{figure}

The long wavelength baseline available in our spectrum of NGC\,7714
makes possible a very accurate determination of the continuum, which
allows sophisticated fits for the dust emission which is responsible
for this continuum.  Figure~\ref{figbbfit} shows a three component fit
to the low-resolution spectrum.  Two components are warm and cool
astronomical silicate/graphite (Si/Gr) dust with an MRN distribution
\citep{mat77} and the optical constants given by \citet{dra84} and
\citet{lao93}.  The third component is astronomical PAHs for which we
use the absorption coefficient of \citet{sie01}. We also allowed for
the possibility of an intervening, absorbing Si/Gr dust screen but
found that $\tau_{9.7\mu m} \leq 0.2$ (consistent with the above
results).  The Si/Gr dust components have temperatures of
approximately 153\,K and 60\,K, and the PAH component has a
temperature of about 340\,K.  PAH emission clearly dominates below
$9\mu$m while the Si/Gr mix fills in the ``PAH trough'' at $10.5\mu$m.
The Si/Gr component then dominates at longer wavelengths.  A warm and
a hot component are clearly present in NGC\,7714.  Due to the absence
of silicate absorption, we have confidence that the continuum is
directly observed, without being obscured by cooler, intervening dust.
This means that we have achieved a direct measurement of the hottest
dust in a prototypical galaxy.  This determination is important
because the dust temperature is often used as a discriminant between
starbursts and AGN.




\acknowledgments

This work is based [in part] on observations made with the Spitzer
Space Telescope, which is operated by the Jet Propulsion Laboratory,
California Institute of Technology under NASA contract 1407. Support
for this work was provided by NASA through Contract Number 1257184
issued by JPL/Caltech.


\end{document}